\documentclass[10pt, conference, compsocconf]{IEEEtran}
\usepackage{cite}

\ifCLASSINFOpdf
   \usepackage[pdftex]{graphicx}
\else
\fi
\usepackage[cmex10]{amsmath}
\usepackage{amssymb}
\usepackage{amsthm}
\usepackage{dsfont}
\usepackage{array}
\usepackage{url}

\usepackage[table,xcdraw]{xcolor}
\usepackage{multirow}
\usepackage{multicol}

\begin{document}
%
\title{Exponential Random Graph Models with Big Networks: Maximum Pseudolikelihood Estimation and the Parametric Bootstrap}


\author{\IEEEauthorblockN{Christian S. Schmid}
\IEEEauthorblockA{Department of Statistics\\
The Pennsylvania State University\\
State College, USA\\
Email: schmid@psu.edu}
\and
\IEEEauthorblockN{Bruce A. Desmarais}
\IEEEauthorblockA{Department of Political Science\\
The Pennsylvania State University\\
State College, USA\\
Email:  bdesmarais@psu.edu}
}


%


\maketitle

\begin{abstract}
With the growth of interest in network data across fields, the Exponential Random Graph Model (ERGM) has emerged as the leading approach to the statistical analysis of network data. ERGM parameter estimation requires the approximation of an intractable normalizing constant. Simulation methods represent the state-of-the-art approach to approximating the normalizing constant, leading to estimation by Monte Carlo maximum likelihood (MCMLE). MCMLE is accurate when a large sample of networks is used to approximate the normalizing constant. However, MCMLE is computationally expensive, and may be prohibitively so if the size of the network is on the order of 1,000 nodes (i.e., one million potential ties) or greater. When the network is large, one option is maximum pseudolikelihood estimation (MPLE). The standard MPLE is simple and fast, but generally underestimates standard errors. We show that a resampling method---the parametric bootstrap---results in accurate coverage probabilities for confidence intervals. We find that bootstrapped MPLE can be run in 1/5th the time of MCMLE. We study the relative performance of MCMLE and MPLE with simulation studies, and illustrate the two different approaches by applying them to a network of bills introduced in the United State Senate. 

\end{abstract}


%
\IEEEpeerreviewmaketitle

\section{Introduction}
The field of network science faces a double-edge sword when it comes to computationally intensive research. First, the availability of digital source data has led the growth in network science to be synonymous with the growth in research on big data.  Faraj et al. (2008, pp. 19--20) \cite{faraj2008electronic} notes that, \begin{quote} ``In recent years the proliferation of advanced Information Technology has not only facilitated the collection of large-scale network data but also increased the availability of large-scale network data analysis tools and techniques. This has led to an explosion of interest in large network research in fields spanning biology, physics, mathematics, and social sciences.''  \end{quote} Second, analytical methods are growing more sophisticated, increasingly involving iterative and/or simulation-based optimization, rather than simple descriptive calculations \cite{snijders2011statistical}.  Closing the gap in terms of the size of the networks to which it is feasible to apply the most sophisticated methods of network modeling requires research into scalable methods of inference. We propose a method of statistical inference for one of the most popular models for networks---the exponential random graph model (ERGM), in which both parameter estimates and confidence intervals are derived, that can require less than half the compute time of currently used methods. 

\section{The Exponential Random Graph Model}
The ERGM is a probabilistic model for networks \cite{chatterjee2013estimating,Wasserman.1996,robins.pattison.2007}. They can be used for link prediction \cite{lu2010supervised}, simulating network adjacency matrices \cite{hackney2006agent}, and testing theories regarding the processes underlying tie formation \cite{goodreau2009birds}. The ERGM was first introduced by Holland and Leinhardt (1981) \cite{holland1981exponential}. However, due to the intractable normalizing constant in the likelihood function of the ERGM, it did not see widespread and complete use until the 2000s, following the development of algorithms and software for efficient simulation-based methods for working with ERGM \cite{snijders2002markov}. Training ERGM using simulation-based methods is computationally expensive, and can still be prohibitively burdensome with data on big networks. Approximate methods of estimation, which are much more feasible with large networks, have existed for some time, but these methods perform poorly when it comes to characterizing the uncertainty in parameter estimates, which is necessary when assessing risk in predictions or simulation, or in hypothesis testing.\\
\indent The ERGM takes the adjacency matrix of an observed network $G^{obs}$, which is a matrix-valued random variable. This means that a network of $N$ nodes can be defined as a adjacency matrix $G=(g_{ij})\in \{0,1\}^{(N \times N)}$, where $g_{ij} \in \{0,1\}$ for all $i,j \in \{1,\dots , N\}$. $g_{ij}=1$ means that there is an edge between actors $i$ and $j$, while $g_{ij}=0$ indicates that these actors are not directly connected. Since the model does not consider loops, one has $g_{ii}=0$ for all $i \in \{1,\dots , N\}$. Furthermore, define
$$ \mathcal{G}(N) := \left\{ G \in \{0,1\}^{(N \times N)}: g_{ij} \in \{0,1\},~g_{ii}=0\right\}$$
as the set of all possible networks on $N$ nodes without loops. Note that the cardinality of set $\mathcal{G}(N)$ is increasing exponentially for every newly included actor, which results in $2^{N(N-1)/2}$ total elements. For even a small number of nodes, the cardinality of $\mathcal{G}(N)$ turns out to be an astronomically large number. For this reason calculating the likelihood function of the ERGM, which requires evaluating a normalizing constant on $\mathcal{G}(N)$  is either extremely time-consuming or with today's technology not achievable. As a consequence, many approximation methods have been provided by the literature, with the most popular method making use of Markov Chain Monte Carlo (MCMC) methods \cite{hunter2012computational}, as we will introduce in the next section.\\  
The probability function for the ERGM is defined as
\begin{equation}
\mathbb{P}_{\theta}(G)=\dfrac{\exp(\theta^T \cdot \Gamma(G))}{\sum_{G^* \in \mathcal{G}(N)} \exp(\theta^T \cdot \Gamma(G^*))} 
\label{ERGM}
\end{equation}
where $\theta \in \mathbb{R}^q$ is a $q-$dimensional vector of parameters, $\Gamma:\mathcal{G}(N) \to \mathbb{R}^q~,~G \mapsto (\Gamma_1(G),\dots,\Gamma_q(G))^T$ is a $q$-dimensional function of different network statistics and $c(\theta):= \sum_{G^* \in \mathcal{G}(N)} \exp(\theta^T \cdot \Gamma(G^*))$ is a normalization constant which ensures that (\ref{ERGM}) defines a probability function on $\mathcal{G}$.
As already mentioned, a specific network $G$ can be considered as a manifestation of a matrix-like random variable, whose probability of occurrence can be modeled with equation (\ref{ERGM}). The generative processes captured by a model (e.g., density, reciprocity, popularity, clustering) are informed by the decision regarding which network statistics (i.e., $\Gamma(\cdot)$) are incorporated (see Snijders et al. \cite{SnijdersTomA.B..2006} for a detailed discussion). The flexibility of the ERGM in capturing virtually any network generative process has led to it being applied broadly across several fields, including (but not limited to) sociology \cite{smith2016ethnic,srivastava2011culture}, economics \cite{lomi2012networks}, political science \cite{Cranmer.2011,lazer2010coevolution}, ecology \cite{dey2014individual,bouskila2015similarity}, and neuroscience \cite{simpson2011exponential,simpson2012exponential}.

\section{Estimation}
As mentioned above calculating $c(\theta)$ is not achievable for most cases with today's technology.  Straightforward application of either maximum likelihood estimation and Bayesian inference are, therefore, not possible. The first method proposed in the literature for estimating ERGM parameters was maximum pseudolikelihood estimation \cite{StraussIkeda1990}. Under maximum pseudolikelihood estimation (MPLE), the joint distribution is replaced by the product over conditional distributions \cite{besag1986statistical}.  
\begin{equation*}
\mathbb{P}_{\theta}(G) \approx \prod_{ij} \mathbb{P}_{\theta}\left(g_{ij}|G_{-ij}\right), 
\end{equation*}
where $G_{-ij}$ is the adjacency matrix, excluding element $ij$. The conditional probability of a tie in ERGM reduces, conveniently, to a logistic regression form given by 
\begin{equation*}
\mathbb{P}_{\theta}\left(g_{ij}=1| G_{-ij} \right) = \text{\\} \text{logit}^{-1}\left( \theta^T \cdot \delta(\Gamma(G))  \right), 
\end{equation*}
where $\delta(\Gamma(G))$ is the ``change statistic'' given by the difference in the network statistics when the $ij$ element is toggled from 0 to 1 (i.e., $\Gamma(G|g_{ij}=1))-\Gamma(G|g_{ij}=0)$), and $\text{logit}^{-1}(x)=1/(1+\exp(-x))$ \cite{goodreau2009birds}. For the ERGM, the pseudolikelihood function can be maximized using logistic regression software, in which the dependent variable is given by the elements of the adjacency matrix, and the covariates are given by the values of the change statistics corresponding to each element of the adjacency matrix. 

Despite the computational efficiency underlying the implementation of MPLE,  existing methods for assessing uncertainty with respect to the MPLE perform poorly (see van Duijn et al. \cite{vanDuijnetal2009}). Estimating the uncertainty in parameter estimates (e.g., standard errors, confidence intervals), is a critical step in using the results from a statistical model. Estimates of uncertainty are used to test hypotheses about parameters, estimate variance (i.e., risk) in model predictions, and estimate effect sizes. Another limitation of MPLE regards the assessment of model fit. With ERGM, model fit is conventionally evaluated by comparing the structure of the observed network to networks simulated from the estimated model \cite{Hunter.2008}. This comparison can include diagnosing the model for a particular form of poor fit that arises with ERGM---model degeneracy \cite{schweinberger2011instability}. Degeneracy is a condition in which the model places nearly all of the probability mass on the completely empty or completely full network. Since MPLE does not require simulating networks, researchers can derive and report results without checking the fit of the model or checking for degeneracy. Of course, the researcher can choose to simulate networks and check model fit with MPLE , but unlike the simulation-based methods of estimation, it is possible to run MPLE without simulating networks.

The contemporary conventional approach to estimating $\theta$, introduced by Snijders \cite{snijders2002markov}, is based on a Markov Chain Monte Carlo (MCMC) approximation of the MLE.\\
This Monte Carlo maximum likelihood method (MCMLE) is based on a more direct attempt to approximate the  log-likelihood function derived from (\ref{ERGM}). The log-likelihood function is not evaluated directly, rather, the log ratio of the likelihood under a proposed value of the parameters $\theta$, and an initial value of the parameters $\theta_0$, is approximated using $L$ networks simulated from the ERGM with parameter values $\theta_0$. The approximation, detailed in Snijders (2002) \cite{snijders2002markov} is given by 
\begin{eqnarray*}
&&\text{loglik}(\theta)-\text{loglik}(\theta_0)\\&=&- \log(c(\theta))+\log(c(\theta_0))\\
                                             &=&- \log \left( \frac{c(\theta)}{c(\theta_0)} \right)\\
                                             &=&- \log \left( \mathbb{E}_{\theta_0}\left[ \exp\left((\theta - \theta_0)^T \cdot \Gamma(Y)\right) \right] \right)\\
                                             &\approx &- \log \left( \frac{1}{L} \cdot \sum_{i=1}^{L}  \exp \left((\theta - \theta_0)^T \cdot \Gamma(G_i) \right) \right).
\end{eqnarray*}
By differentiating this equation on both sides with respect to $\theta$ one gets an approximate score function:
\begin{equation}
s(\theta) \approx -\frac{\partial}{\partial \theta} \log \left( \frac{1}{L} \cdot \sum_{i=1}^{L}  \exp \left((\theta - \theta_0)^T \cdot \Gamma(G_i) \right) \right)
\label{score}
\end{equation}
This approximate score function now can be used as usual, i.e., it can be iteratively approximately optimized with the \textit{Newton-Raphson algorithm} or by \textit{Fisher scoring}. 

MCMC methods are used to simulate the $L$ networks. As we demonstrate below, the MCMLE grows more accurate as $L$ increases. Indeed, MCMLE approaches the MLE as the number of networks simulated goes to infinity. Snijders (2002) \cite{snijders2002markov} provides a Metropolis Hastings algorithm to simulate networks:
Choose a matrix $G^{(0)} \in \mathcal{G}(N)$ to start with (e.g., the observed network). For $k \in \{0,...,L-1\}$ recursively proceed as follows:\\
\begin{enumerate}
\item Randomly choose an edge $(i,j)$ where $i \neq j$ from $G^{(k)}$
\item Compute the value
$$\pi := \dfrac{\mathbb{P}_{\theta}(Y_{ij} \neq g_{ij}^{(k)} | Y_{ij}^c=G_{ij}^c)}{\mathbb{P}_{\theta}(Y_{ij}= g_{ij}^{(k)} | Y_{ij}^c=G_{ij}^c)}$$
\item Fix $\delta:= \min\{1, \pi\}$ and draw a random number $Z$ from Bin$(1, \delta)$. If
\begin{itemize}
\item $Z=0$, let $G^{(k+1)} := G^{(k)}$ 
\item $Z=1$, define $G^{(k+1)}$ via
$$g_{pq}^{(k+1)}=\begin{cases}
1-g_{pq}^{(k)}& \text{if}~ (p,q)=(i,j) \\
g_{pq}^{(k)} &\text{if}~ (p,q) \neq (i,j) 
\end{cases}$$
\end{itemize}
\item Start at step 1 with $G^{(k+1)}$.
\end{enumerate}
\vspace{0.3cm}
The depicted algorithm provides a sequence of random networks $G_(0),\dots,G_(L)$ via a Metropolis-Hastings sampler \cite{chib1995understanding}. Since the original matrix was chosen randomly and the first simulated networks are very dependent on the chosen matrix (only one edge is changed per iteration), usually
the first $B$ networks, where $N << B << L$, are discarded as the so called Burn-In.

\section{Efficiency of MPLE and MCMLE}
As mentioned in the previous section the MPLE approaches the MLE as the size of the networks increase and as a consequence, is a consistent estimator (see Lindsay \cite{Lindsay1988}, Strauss and Ikeda \cite{StraussIkeda1990}, Hyvarinen \cite{Hyvarinen2006}, Desmarais and Cranmer \cite{Desmarais.2012,desmarais2010consistent}). This implies that for an increasing number of nodes, the MPLE converges in probability to the MLE, meaning that for large enough networks the MPLE performs as well as or better than MCMLE, and requires less compute time. At this point we want to mention that we are familiar with the work of Shalizi and Rinaldo \cite{shalizi2013}, arguing that consistency is not given in the ERGM framework. They prove that one cannot run an ERGM on a sub-network in order to make inferences about the full network.  The way we use the term \textit{consistency} in this paper is different and aligns with the way consistency is defined by Lindsay \cite{Lindsay1988}, i.e. instead of considering sub-networks that converge to the full size network, we argue that both, the MLE as well as the MPLE, approach the true coefficient values as the size of networks generally increases.

\begin{figure}[!t]
\centering
\includegraphics[width=3.6in]{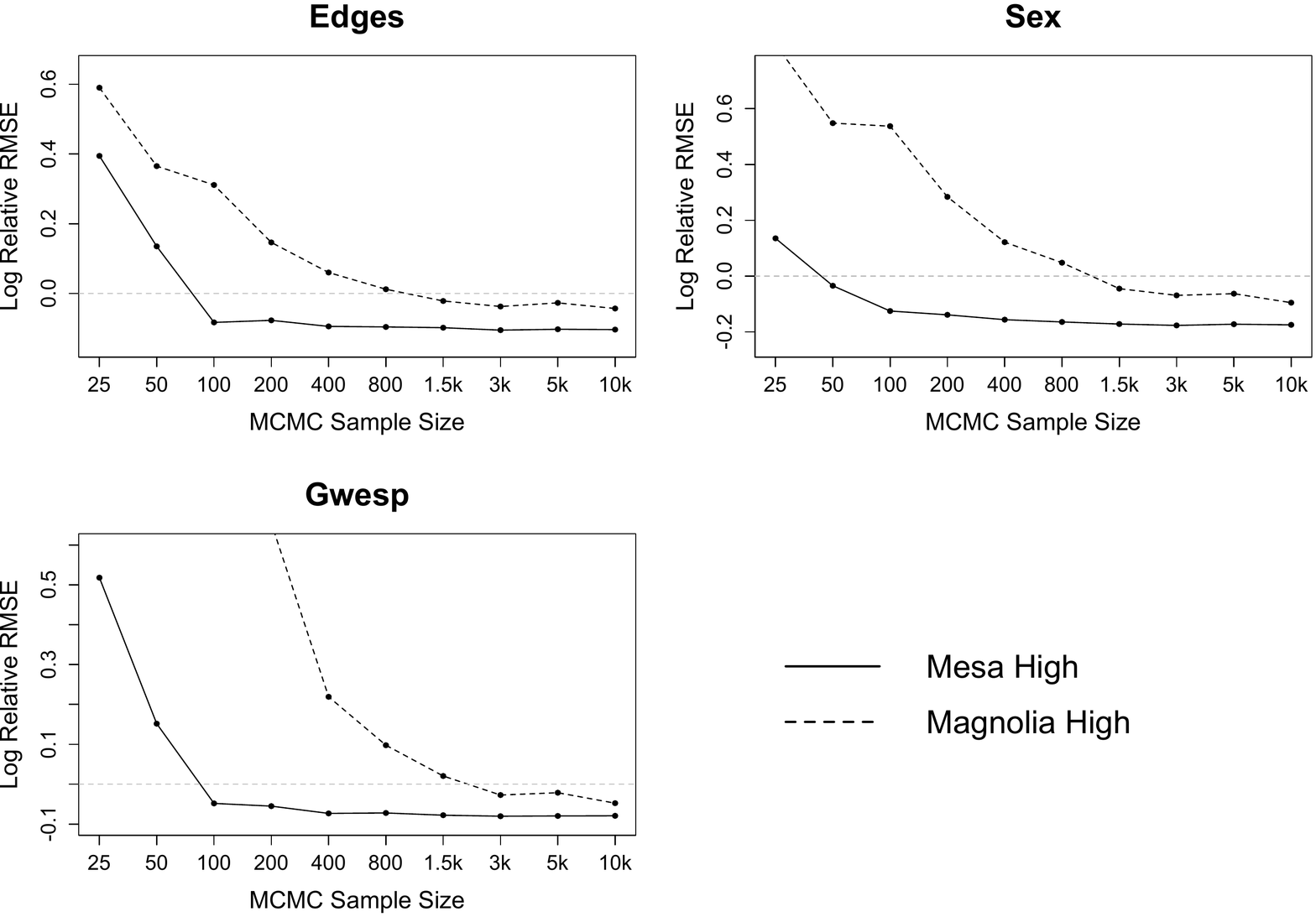}
\caption{The log of the ratio of the RMSE for the MCMLE to the MPLE for different sample sizes and two different networks, Faux Mesa High and Faux Magnolia High}
\label{rmse}
\end{figure}

To illustrate the relative efficiency of MPLE and MCMLE we run a simulation study. Desmarais and Cranmer \cite{Desmarais.2012} show the MPLE outperforms the MCMLE if the number of simulated networks used to approximate the likelihood in MCMLE is not large enough. It is even more remarkable that the number of simulated networks needed for the MCMLE, in order to surpass the MPLE increases as the size of the network (i.e., the number of nodes in the network) increases. This means that, for very large networks, it becomes difficult to determine the number of simulated networks required for the MCMLE to outperform the MPLE. In other words, the larger the network, the more computationally intensive it becomes to use MCMLE in a way that out-performs MPLE.\\
\indent To demonstrate this disadvantage of the MCMLE we conduct a simulation study using Goodreau's \cite{HunterDavidR..2008} Faux Mesa High School data, which represents a simulation of an in-school friendship network among 203 students as well as the Faux Magnolia High School data, representing an in-school friendship network among 1451 students. The data for both networks originates from Resnick et al. \cite{Resnicketal1997}.\\
\indent For both networks, we first calculate the MCMLE and treat the estimated coefficients as the network's true values $\theta$. Then, we take the same parametrization, using the number of edges, the nodal attribute for gender, and the geometrically weighted edgewise shared partners (gwesp) distribution (see Hunter \cite{Hunter.2006}) where we fix the decay parameter $\lambda$ at $0.25$.
The number of edges is defined as
$$\Gamma_{edges}: \mathcal{G}(N) \to \mathbb{R}~~~, ~~~ G \to \sum_{i<j}^N g_{ij}$$
In order to include nodal covariates into the ERGM, the vector of nodal attribute is expanded into a matrix $C$, which has the same dimensions as $G$. The first row of matrix $C$ consists of the first actor's attribute, repeated $N$ times. The second row of matrix $C$, consists of the second actor's attribute, repeated $N$ times, and so on. Then, the statistic for a nodal covariate is defined as
$$\Gamma_{nodal}: \mathcal{G}(N) \to \mathbb{R}~~,~~G \mapsto \sum_{i<j}^{N}g_{ij}c_{ij}$$
The GWESP statistic is given as
$$\Gamma_{gwesp}(G,\lambda):= \lambda\sum_{j=1}^{N-1}\Biggr(1- \biggr(1-\cfrac{1}{\lambda}\biggl)^j\Biggl)\Gamma_{esp(k)}(G)$$
where 
$$
\Gamma_{esp}: \mathcal{G}(N) \to \mathbb{R}~~,~~G \mapsto \sum_{i<j}^{N}\mathds{1}_k\biggl(\sum_{k=1}^N g_{ij}g_{im}g_{jm}\biggr).
$$
$\Gamma_{esp}(G,k)$ counts the number of nodal pairs $(i,j)$, which share exactly $k$ neighbors. This statistic is used to model the tendency towards triangles and clustering in a network.\\
\indent We simulate $m=500$ new networks using the 'true' coefficients and estimate the MPLE as well as the MCMLE of these simulated networks. For every single simulated network the MCMLE calculation is being repeated several times for $25$ to $10,000$ simulated networks used in the likelihood approximation.
Based on these results, we compute the root mean square error, which is a measure of the accuracy of an estimator, combining both the bias and the variance. Mathematically written, the RMSE for an estimator $\hat{\theta}$ is defined as 
$$RMSE = \sqrt{\sum_{i=1}^{m}(\theta - \hat{\theta}_i)^2}$$
implying that the smaller the RMSE, the more accurate is the estimator. Since the MCMLE has higher efficiency and converges to the MLE, the RMSE decreases as the number of simulated networks used for the likelihood approximation increases. On the other hand, the RMSE of the MPLE is a constant value since no network simulations are required. In order to compare the RMSE of the two estimation techniques, we take the log of the ratio of the MCMLE to the MPLE. As a result, a negative value indicates a better MCMLE performance, while a positive value indicates a better MPLE performance.\\
\indent Figure \ref{rmse} visualizes the results of the simulation study. The solid line illustrates the results of the log relative RMSE of the Faux Mesa High network, while the dashed line illustrates the corresponding results of the Faux Magnolia High network. 
The plots support the fact that larger networks require a larger sample size of simulated networks for the MCMLE to outperform the MPLE. While the fairly small Faux Mesa High network only requires a sample size of about $50-100$ networks, the larger Faux Magnolia High network requires a sample size of at least 1,500 networks for the MCMLE to surpass the MPLE.  For especially large networks (e.g., social media data) the sample size has to be set in order to justify the approximately exact, but computationally expensive and potentially prohibitive MCMLE method.

\section{Bootstrapped MPLE}
As discussed in the previous section, the MPLE converges to the MLE as the size of the network increases. Moreover, the MPLE is able to outperform the MCMLE if the sample size used in MCMLE is not large enough. The main reason why the MCMLE is still widely preferred is that, in contrast to the MPLE, it does not underestimate the standard errors (van Duijn et al. \cite{vanDuijnetal2009}). By the definition of the ERGM it is obvious that this model is an exponential family distribution where $\theta$ is the natural parameter and $\Gamma (G)$ is the sufficient statistic. For exponential family distributions, it is known that the sampling distribution of the MLE is multivariate normal with mean vector equal to the MLE and a covariance matrix equal to the inverse of the negative Hessian matrix $[-H]^{-1}$ of the likelihood function at the MLE. The problem with the MPLE is that calculating $[-H]^{-1}$ by the pseudolikelihood function will underestimate the variance of the MPLE \cite{vanDuijnetal2009}, resulting in an underestimate of the width of the confidence intervals. van Duijn et al. show that constructing 95\% MPLE confidence intervals can result in intervals that comprise the true value in less than 75\% instead of the nominal 95\%. In this paper, we are going to refer to the MPLE confidence intervals as \textit{logistic regression confidence intervals} simply because the MPLE is calculated using logistic regression methods that also use the inverse of the negative Hessian matrix as an estimate for the covariance matrix. \\
\indent Since the MPLE has the advantage of being approximately exact and computationally inexpensive, but has the disadvantage of underestimating corresponding confidence intervals, we apply a technique referred to as bootstrap resampling \cite{efron1982jackknife}. Bootstrap resampling refers to constructing a sampling distribution for the parameter estimate by resampling the data with replacement, and re-estimating the model on the resampled data. Under non-parametric bootstrap resampling, the data are resampled directly from the dataset. Under the parametric boostrap, the data is resampled from the estimated model. The idea of using boostrap resampling with MPLE for ERGM was first introduced by Desmarais and Cranmer \cite{Desmarais.2012} and provides a consistent estimate of MPLE confidence intervals. Desmarais and Cranmer argue that the MPLE is a multivariate \textit{M}-estimator (see Huber \cite{Huber1981}), a special class of robust estimators, meaning that bootstrap resampling consistently estimates the confidence intervals of the MPLE. \\
\indent However, the methods introduced by Desmarais and Cranmer \cite{Desmarais.2012} only applied to cases in which the researcher had a large sample of networks (e.g., a time series of networks), since the method they proposed was a non-parametric bootstrap. The non-parametric bootstrap cannot be applied when there is just a single network under study. 

For the case in which there is just a single network to be studied, which is indeed the most common case in the literature, we propose the use of a parametric bootstrap. Under the parametric bootstrap, the sampling distribution of the MPLE is derived by re-estimating the MPLE on a sample of networks simulated from the MPLE estimated on the observed network. We verify the consistency of the bootstrapped MPLE by conducting a simulation study on the same two networks with the same parametrization as in the previous chapter: The Faux Mesa High friendship network and the Faux Magnolia High friendship network. 

For the simulation study, we determine the MPLE for the model and treat these estimates as the networks' 'true' parameter values. We then use these parameter values to simulate a sample of $1000$ networks from the distribution of $G$. For each of the $1000$ networks, we calculate 95\% confidence intervals based on the MCMLE and the logistic regression and examine whether the 'true' parameter values lie in these intervals. In addition, we determine the bootstrapped MPLE confidence intervals by sampling $500$ networks for each of the originally sampled $1000$ networks, by using the respective MPLE as parameter values. For every newly sampled network, we again determine the MPLE and then take the $2.5$th and $97.5$th percentile of the $500$ MPLE estimates to obtain  95\% bootstrap confidence intervals. Similar as for the MCMLE and the logistic regression, we verify whether the 'true' parameter value can be found in the bootstrapped confidence interval. 
\begin{figure}[!t]
\centering
\includegraphics[width=3.5in]{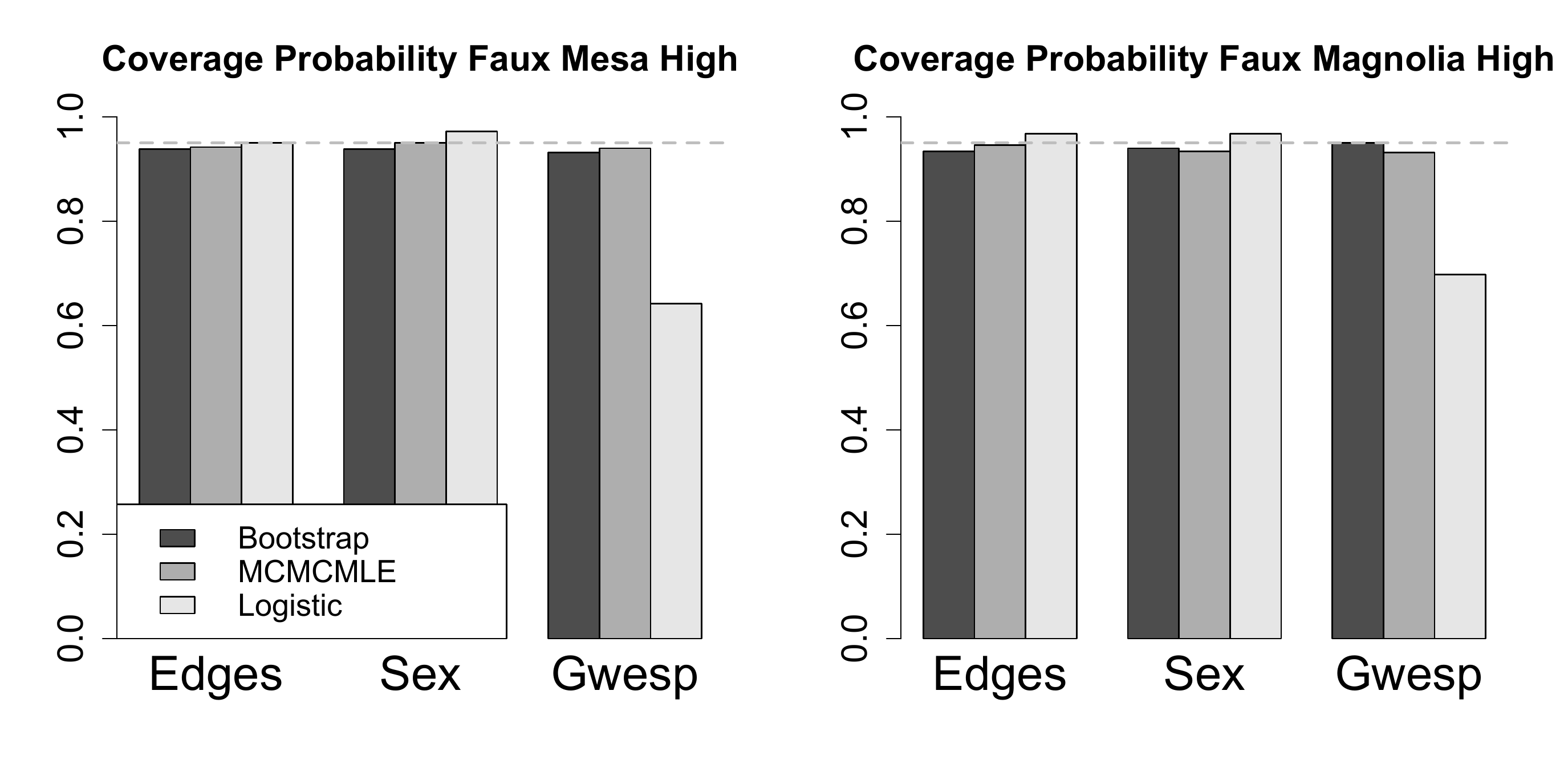}
\caption{The Coverage Probability results of the Faux Mesa High network (left) and of the Faux Magnolia High network (right) for bootstrapped MPLE, MCMLE and logistic regression}
\label{coverage}
\end{figure}

\begin{figure}[!t]
\centering
\includegraphics[width=3.5in, height=1.75in]{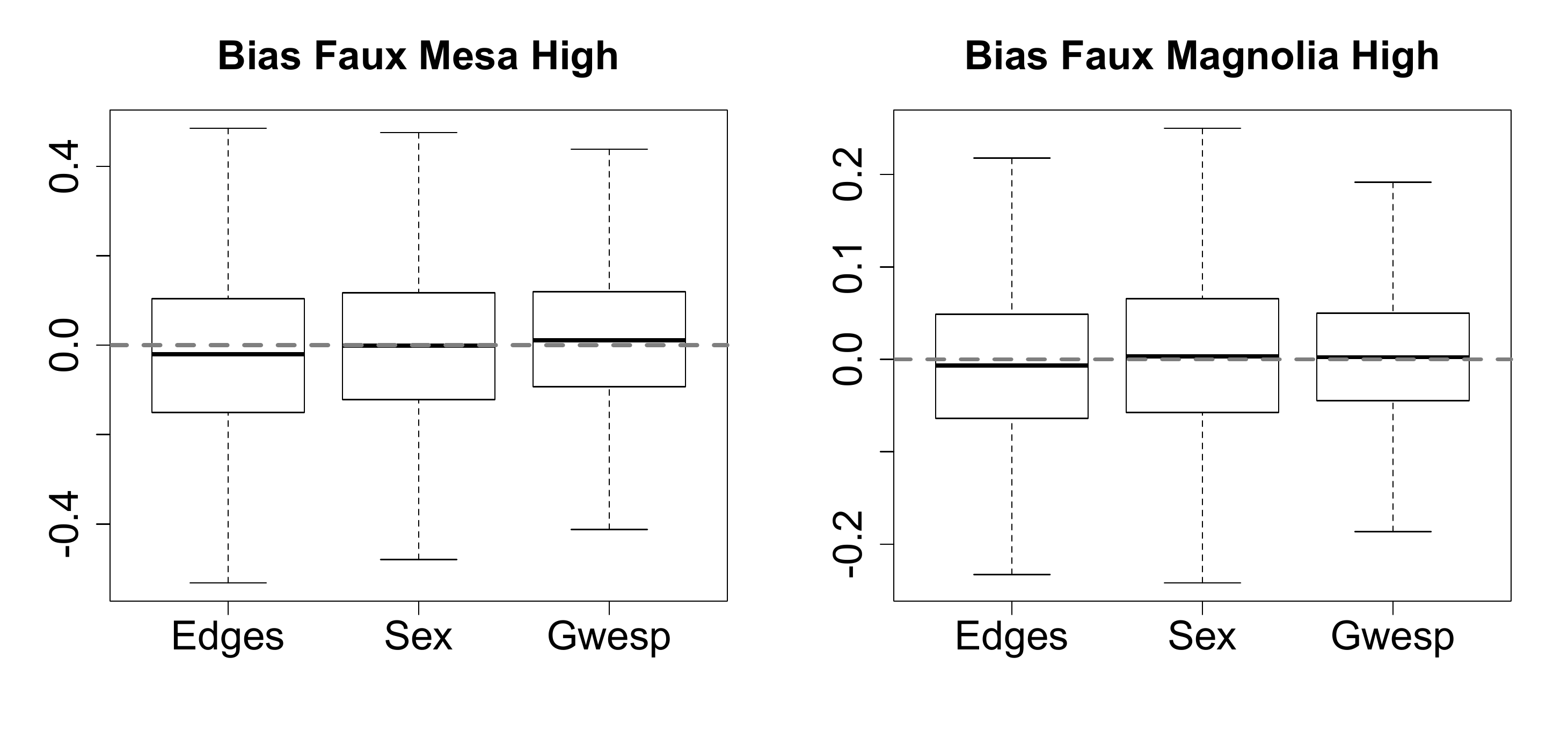}
\caption{The boxplots visualize the bias $(\hat{\theta}-\theta)$ over the 500 iterations for the Faux Mesa High network (left) and the Faux Magnolia High network (right)}
\label{bias}
\end{figure}
Figure \ref{coverage} visualizes the coverage percentages for each of the three methods for both networks. The dashed line is set at $0.95$ and represents the optimal value. It is evident that the bootstrapped MPLE performed equally well as the MCMLE, achieving results that obtain the true parameter values in approximately 95\% of the cases. Additionally, a difference in the results between the smaller Faux Mesa High network and larger Faux Magnolia High network is not identifiable. Similar to the results of van Duijn et al. \cite{vanDuijnetal2009} our results for the logistic regression differ distinctively from the anticipated 95\%, confirming that the MPLE underestimates the variance of its estimates.
Figure \ref{bias} illustrates the bias between the 'true' network coefficients $\theta$ and the MPLE estimates. We can abstract from this figure that the median MPLE estimates concur with the networks 'true' coefficients. It is especially worthwhile to mention that the bias of the larger Faux Magnolia High network is smaller than the bias of the Faux Mesa High network, supporting the fact that the MPLE converges to the MLE as the network size increases.\\
\indent This simulation study shows that bootstrapped MPLE is able to overcome the main disadvantage of the MPLE by retaining the validity of confidence intervals. In this simulation study we demonstrate that the parametric bootstrap can be used in combination with MPLE to provide a method of estimation that is both computationally efficient and provides valid estimates of model uncertainty.

\section{Cosponsorship Network Data}
\begin{table*}[!t]
\renewcommand{\arraystretch}{1.3}
\label{co_results}
\centering
\begin{tabular}{|
>{\columncolor[HTML]{EFEFEF}}l |l|l|l|l|l|l|}
\hline
\cellcolor[HTML]{EFEFEF}                   & \multicolumn{2}{c|}{\cellcolor[HTML]{EFEFEF}MCMLE}                   & \multicolumn{2}{c|}{\cellcolor[HTML]{EFEFEF}Logistic Regression}       & \multicolumn{2}{c|}{\cellcolor[HTML]{EFEFEF}bootstrapped MPLE}     \\ \cline{2-7} 
\multirow{-2}{*}{\cellcolor[HTML]{EFEFEF}} & \cellcolor[HTML]{EFEFEF}Estimate & \cellcolor[HTML]{EFEFEF}St. Error & \cellcolor[HTML]{EFEFEF}Estimate & \cellcolor[HTML]{EFEFEF}St. Error & \cellcolor[HTML]{EFEFEF}Lower Bound & \cellcolor[HTML]{EFEFEF}Upper Bound \\ \hline
Edges                                      & -5.884 & 0.065                             & -5.869                       & 0.015                                    & -6.007 & -5.751  \\ \hline
Sponsor Party                          & 1.440                            & 0.015                             &   1.440                               & 0.015                                   & 1.411 & 1.467  \\ \hline
Alternating k-star                              & 0.124                            & 0.064                             &     0.108                             & 0.006                                   & -0.011 & 0.2379  \\ \hline
\end{tabular}\vspace{.2cm}
\caption{Estimation results for the Cosponsorship network using MCMLE, logistic regression and bootstrapped MPLE}
\end{table*}
To illustrate the performance of MCMLE relative to that of the bootstrapped MPLE we apply both approaches to the data on cosponsorship of bills in the U.S. House of Representatives for the 108th Congress (2003--2004), developed by Fowler (2006) \cite{Fowler2006a} \cite{Fowler2006b}. The cosponsorship network consists of 2,635 nodes, which we define as pieces of legislation (i.e., bills), considered by the Senate during the 108th Congress. Note that this formulation of the cosponsorship network differs from past research, which has defined the nodes as the individual legislators. Because there are more bills than legislators, studying bills as nodes provides a more disaggregated look at the network than is offered through studying the network of legislators. In this undirected network bills are tied together based on the similarity of the sets of legislators who cosponsor them. Specifically, we include an edge between bills $i$ and $j$ if the correlation coefficient between the indicator vectors indicating whether $i$ and $j$ were sponsored each legislator is greater than a random uniform draw. This results in an undirected network with $28060$ edges.

We build an ERGM specification that extends the work of  Zhang et al\cite{zhang2008community} in exploring the structure of cosponsorship ties. They find that congressional cosponsorship is primarily characterized by intra-party ties---among Republicans and among Democrats, but few cross-party ties. We test for this party-based clustering (i.e., homophily) in our ERGM. This is done through a term that accounts for the party of the senators who sponsored the two bills in the pair. The party homophily term is defined as $$ \Gamma(G,X) = \sum_{i < j} g_{ij}x_{ij},$$ where X is an indicator matrix that assumes the value 0 if $i$ and $j$ were sponsored by legislators from different political parties and 1 if they were sponsored by legislators from the same party.  $\Gamma(G,X)$ measures the number of intra-party ties in the network. A positive parameter value for this statistic indicates that ties tend to be formed between bills sponsored by the same political party.

We extend the homophily-based model to account for a network dynamic that is commonly found in the study of networks---that of popularity or preferential attachment \cite{barabasi1999emergence}. Preferential attachment is the tendency for new ties to be formed with nodes who already have many ties (i.e., popular nodes).  The alternating k-star statistic was introduced by Snijders et al. \cite{SnijdersTomA.B..2006} and modified by Hunter and Handcock  \cite{Hunter.2006}. A positive parameter estimate associated with the alternating k-star statistic indicates that tie formation follows a form of preferential attachment \cite{SnijdersTomA.B..2006}. This could arise in a network of bill-to-bill ties if the majority party in power was particularly disciplined at rallying its partisans to pile on to the bills that its members proposes, thus producing a large set of very similar bills. The alternating k-star statistic adds one network statistic to the model equal to a weighted alternating sequence of k-star statistics with weight parameter $\lambda$ and is a way to include a networks entire degree distribution as a network statistic. In this model we fix the weight paramter $\lambda=0.4975$. 
Snijders et al. \cite{SnijdersTomA.B..2006} introduced an approach involving $k$-star statistics $S_1(G), \dots , S_{N-1}(G)$, where $S_k(G)$ denotes the number of $k$-stars in the network, $k \in \{1, \dots , N-1\}$. For simplicity, let us define
$$S_k(G):=\Gamma_{star(k)}(G) $$
where
$$\Gamma_{star(k)}: \mathcal{G}(N) \to \mathbb{R}~~~, ~~~ G \to \sum_{i,j,k}^N g_{ij}g_{ik}$$
Note that in every network $S_1(G)=\Gamma_{edges}(G)$, i.e., $S_1(G)$ is equal to the number of edges in the network.
On this basis, Snijders introduces the \textit{alternating k-star statistics}
\begin{equation*}
\begin{split}
\mathfrak{S}(G,\lambda):= & \sum_{k=2}^{N-1}(-\cfrac{1}{\lambda})^{k-2}S_k(G)\\= & ~   S_2(G)-\cfrac{S_3(G)}{\lambda}+ \dots + (-1)^{N-3}\cfrac{S_{N-1}(G)}{\lambda^{N-3}}
\end{split}
\end{equation*} 
Models with this statistic and a fixed decay parameter turn out to be standard ERGMs and Hunter and Handcock \cite{Hunter.2006} succeeded in proving that one can also rewrite alternating k-stars as a function of a network's degree distribution
\begin{equation}
\mathfrak{S}(G,\lambda)= \lambda\Biggr(\lambda\sum_{j=1}^{N-1}\Biggr( 1-\biggr(1-\cfrac{1}{\lambda}\biggl)^j\Biggl)D_j(G) + 2S_1(G)\Biggl)
\label{alto}
\end{equation}
where $D_j(G):= \Gamma_{deg(j)}(G)$ is the number of nodes with a degree of $j$. 
In the next step, we define the \textit{geometrically weighted degree} (gwd) statistic as the first summand of (\ref{alto})
\begin{equation}
\Gamma_{gwd}(G,\lambda):= \lambda\sum_{j=1}^{N-1}\Biggr(1- \biggr(1-\cfrac{1}{\lambda}\biggl)^j\Biggl)D_j(G)
\end{equation}
At this point it also becomes obvious where the \textit{geometrically} comes from. It simply refers to the geometric sequence $(1-\frac{1}{\lambda})^j$ which appears in these statistics.\\
\indent We estimate the ERGM using MCMLE and the bootstrapped MPLE. The MCMLE requires a sample size of at least $1000$ networks to converge. The bootstrapped MPLE was estimated by using 500 simulated networks. As we described in the section {\it Estimation}, only one edge at a time is changed when simulating networks. For better comparison, we chose the same Burn-In ($300,000$ MH-steps) and the same number of iterations ($30,000$ MH-steps) for sampling networks. The results can be found in table \ref{co_results}.\\
\indent It is interesting to see that the standard error calculated by the logistic regression approach is much smaller than the standard error of the MCMLE for the alternating k-star statistic, which leads to inaccurate confidence intervals as shown in figure \ref{coverage}. The MPLE estimate is equivalent to the logistic regression estimate, but the bootstrap confidence intervals, especially for the alternating k-star statistic, are much wider than would be calculated using the logistic regression standard errors. An estimate is generally considered statistically different from zero (i.e., statistically significant) if the confidence interval does not contain zero, or if the ratio of the estimate to the standard error exceeds 1.96 in magnitude. This cosponsorship network example perfectly illustrates the inferential problems that can arise with the conventional logistic regression standard errors when using MPLE. All of the parameter estimates are statistically significant according to the logistic regression estimates. However, the alternating k-star statistic is not significant according to either the MCMLE or the bootstrapped MPLE.

\section{Parallel Computing with MPLE}
\begin{figure}[!t]
\centering
\includegraphics[width=3.5in, height=2.7in]{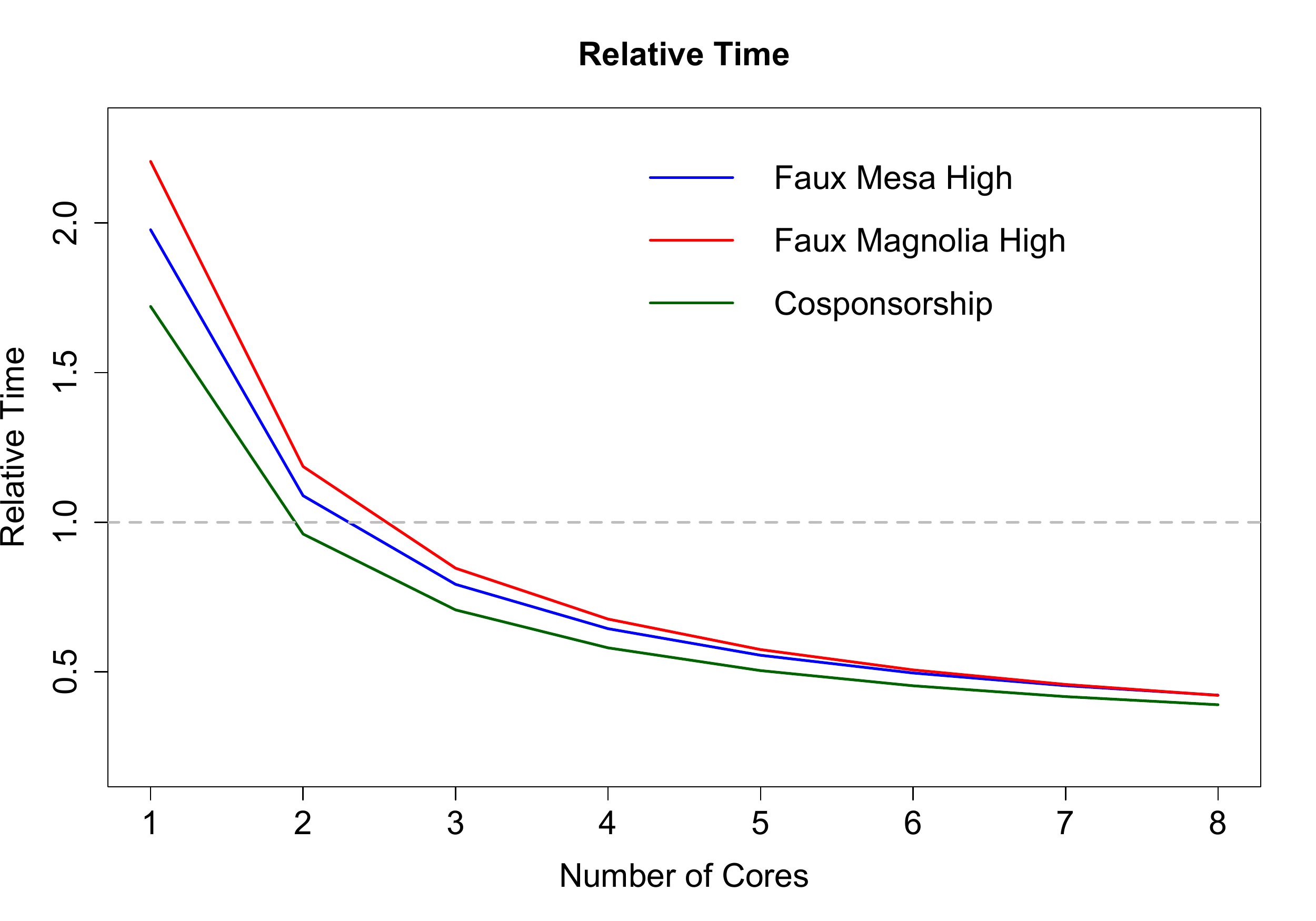}
\caption{The y-axis gives the ratio of the bootstrapped MPLE time to that of the MCMLE time. Values below 1 indicate that the bootstrapped MPLE requires a shorter computing time.}
\label{comptime}
\end{figure}

\begin{figure}[!t]
\centering
\includegraphics[width=3.5in, height=2.7in]{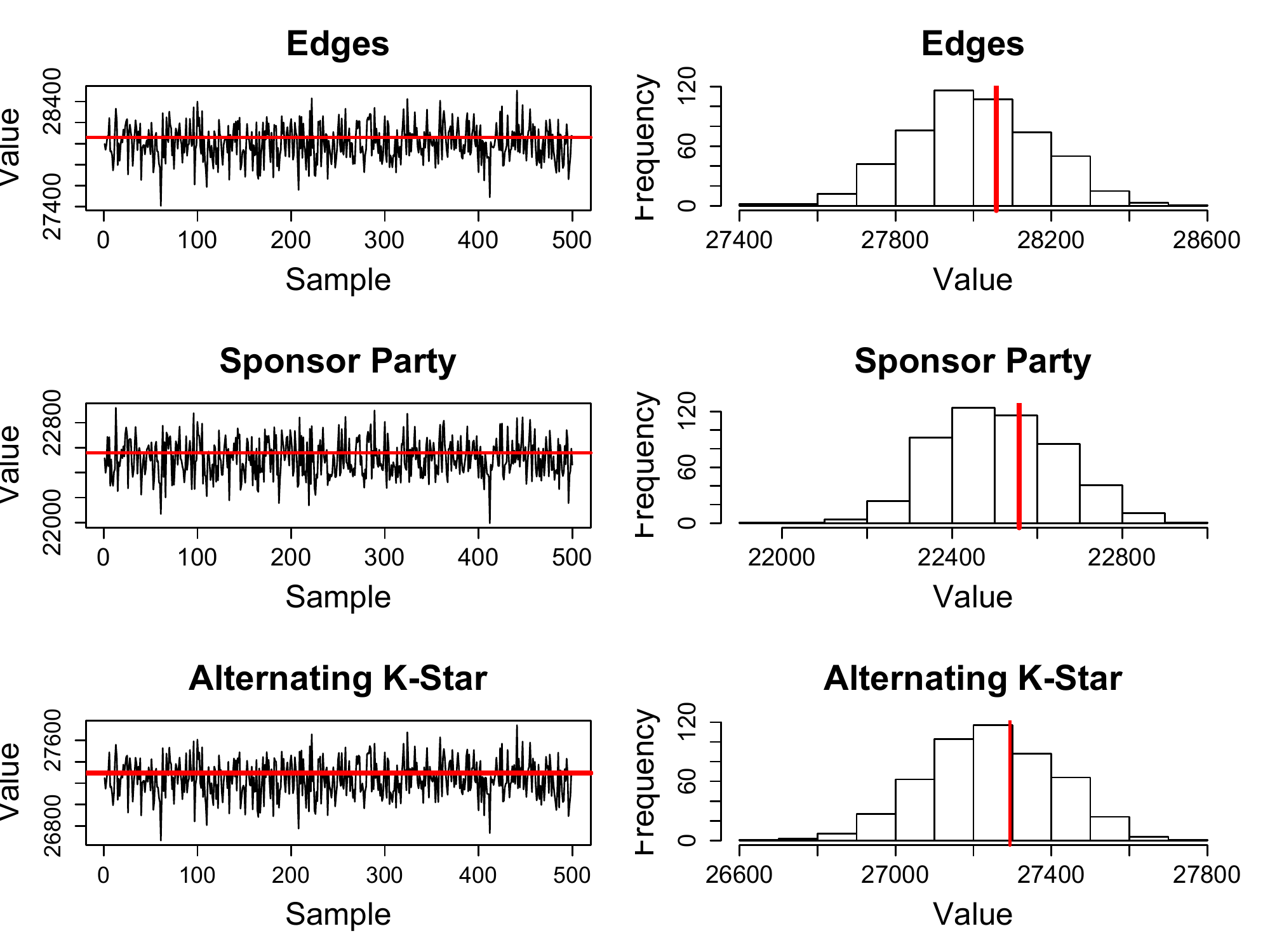}
\caption{Network statistics of the 500 boootstrap samples for the cosponsorship network. The thick line in both, the traceplots and the histograms, represents the network statistics of the observed network. }
\label{diagnostics}
\end{figure}

The bootstrapped MPLE is not only simple and fast, it is highly parallel. Once the networks on which to estimate the bootstrap replicates are simulated, each re-estimate can be run in parallel. By using multiple cores, the computing time for estimating bootstrapped MPLE confidence intervals can be reduced substantially. Figure \ref{comptime} illustrates the relative computing time of the bootstrapped MPLE using 500 simulated networks and the MCMLE for the three networks Faux Mesa High (205 nodes), Faux Magnolia High (1461 nodes) and Cosponsorship (2635 nodes) for an increasing number of computing cores. For the small network we simulate $2000$ networks using a MCMC interval of $2000$ steps, for the medium network we simulate $8000$ networks using a MCMC interval of $5000$ steps and for the large network we simulate $10000$ networks using $30,000$ MCMC steps in order to approximate the likelihood appropriately. The chosen sample sizes and MCMC steps are necessary to guarantee a good model fit. The small network took $14$ seconds, the medium network took $123$ seconds and the large network took $986$ seconds to run. We define the simulation time of the bootstrapped MPLE as a function of the number of available computing cores x:
\begin{equation*}
\begin{split}
\text{boostrapped MPLE time} = ~~~~~~~~~~~~~~~~~~~~~~~~~~~~~~~~~ \\ \text{network simulation time} + \cfrac{500 \cdot \text{MPLE estimation time}}{x}
\end{split}
\end{equation*}     
Based on this, we define the relative computing time as
$$\text{relative computing time}= \cfrac{\text{bootstrapped MPLE time}}{\text{MCMLE time}}$$
This means that a relative computing time greater than 1 indicates that the MCMLE computing time is shorter, while a relative computing time smaller than 1 indicates that the bootstrapped MPLE provides faster results. \\
Figure \ref{comptime} demonstrates that all three networks only require three cores for the bootstrapped MPLE to outperform the computing time of the MCMLE and that the computing time can further be reduced if more computing cores are available. If exactly 500 computing cores are being used the ratio of the bootstrapped MPLE time to the MCMLE time levels off at $0.20$ for the small and large network and $0.17$ for the medium network, meaning that the computing time can be quintupled using the bootstrapped MPLE. This figure also depicts that larger network in general require a longer computation time and will benefit more if the bootstrapped MPLE is used.\\ 
Of course the actual computing time for a network always depends on the statistics that are included in the network, but in general larger networks require longer computation times, since a larger MCMC sample size is required and more MC steps are necessary to simulate a new network that does not overly depend on the previous sample. This makes the bootstrapped MPLE a beneficial alternative, especially for very large networks. \\
\indent One of the major disadvantages of MPLE over MCMLE is that degeneracy is not assessed while the model is being estimated.  The bootstrapped MPLE, however, allows assessing degenerate models as well since the method requires simulating from the estimated parameters. In order to verify whether a model is degenerate or not, one can take a look at density and trace plots as visualized in figure \ref{diagnostics}. The trace plots on the left side depict the the attained values via MCMC simulated networks for every single statistic included into the model, centered on the statistic values of the observed network. The plots on the right side visualize the empirical density function of the respective statistic, based on the simulated networks (Hunter and Handcock \cite{Hunter.2006}). For a non-degenerated model the empirical density function should be approximately symmetrical around zero for every included centered statistic, since this corresponds with the expected value of a centered statistic. 
Otherwise, the values of the simulated networks systematically differ from the corresponding statistics in the observed network, making it unreasonable to assume that the simulated networks originate from the same distribution as the observed network. Furthermore, the trajectories in the trace plot should not indicate a dependence structure. This would be a signal that the constructed stochastic process violates the Markov properties.

\section{Conclusion}
In the past years the ERGM grew in popularity in many different fields as a flexible and powerful means of building probabilistic models for networks. With this popularity the size of the considered networks has grown. As the size of the network increases, it becomes unclear how many simulated networks are necessary for the conventional MCMLE method to perform better than the MPLE. Furthermore, as an increasing number of simulated networks is necessary for the MCMLE, the computing time rapidly grows. For this reason it is essential to develop different methods that provide faster estimation than the MCMLE, but still lead to reliable results.\\
In this paper we introduced the bootstrapped MPLE as an alternative method of statistical inference for ERGMs and compared the performance to the commonly applied MCMLE. Based on a simulation study we demonstrated that the larger the size of a network is the larger the MCMC sample size has to be in order for the MCMLE to outperform the fast and simple MPLE. However, the big disadvantage of the MPLE is that, even though it is an approximately exact estimator, it underestimates the standard error. For this reason, we propose a parametric bootstrap method of evaluating model uncertainty. On the basis of another simulation study on two different networks, we demonstrate that the bootstrapped MPLE covers the true coefficients just as well as the MCMLE, while the simple MPLE performs clearly poorer. This means that the bootstrapped MPLE combines the advantages of both methods, the MPLE and the MCMLE, because it is still simple and fast, and provides approximately exact results, but also accurately estimates model uncertainty. We conclude that the bootstrapped MPLE should be regarded as an alternative to the MCMLE. It also has the advantage of being parallel, which leads to a rapid speed-up of the calculation if multiple computing cores are used.

\section{Acknowledgements}

This work was supported in part by NSF grants SES-1558661, SES-1637089, SES-1619644, and CISE-1320219. Any opinions, findings, and conclusions or recommendations are those of the authors and do not necessarily reflect those of the sponsors.

\bibliography{bib} 
\bibliographystyle{plain}

\end{document}